\documentclass[prd,nofootinbib,showpacs]{revtex4}
\newcommand{\be}{\begin{equation}}\newcommand{\ee}{\end{equation}}
\newcommand{\bea}{\begin{eqnarray}}\newcommand{\eea}{\end{eqnarray}}
\newcommand{\brr}{\begin{array}}\newcommand{\err}{\end{array}}
\newcommand{\bit}{\begin{itemize}}\newcommand{\eit}{\end{itemize}}
\newcommand{\ben}{\begin{enumerate}}\newcommand{\een}{\end{enumerate}}

\def\lab{\label}\def\lan{\langle}
\def\lf{\left}
\def\non{\nonumber}\def\pa{\partial}
\def\ran{\rangle}\def\rar{\to}
\def\ri{\right}
\def\al{\alpha}\def\bt{\beta}\def\de{\delta}
\def\ep{\epsilon}\def\te{\theta}\def\Te{\Theta}
\def\si{\sigma}
\def\om{\omega}\def\AB{{_{A,B}}}
\newcommand{\mlab}[1]{\label{#1}}
\def\AB{{_{A,B}}}\def\mass{{_{1,2}}}
\def\1{{_{1}}}\def\2{{_{2}}}

\begin{document}

\title{Mixing and oscillations of neutral particles in Quantum Field Theory}

\vspace{5mm}

\author{Massimo Blasone }
\affiliation{The Blackett Laboratory, Imperial College London, London SW7
2AZ, U.K.}
\affiliation{Dipartimento di
Fisica and INFN, Universit{\`a} di Salerno, 84100 Salerno, Italy }
\author{Jonathan S. Palmer}
\affiliation{Mathematical Institute, 24-29 St Giles', Oxford OX1 3LB, U.K.}


\begin{abstract}
We study the mixing of neutral particles in Quantum Field Theory: neutral
boson field and Majorana field are treated in the case of mixing among
two generations. We derive the orthogonality of flavor and mass representations
and show how to consistently calculate oscillation formulas, 
which agree with previous
results for charged fields and exhibit corrections with respect to the
usual quantum mechanical expressions.
\end{abstract}


\maketitle

\section{Introduction}

The study of mixing of fields of different masses in the context of
Quantum Field Theory (QFT) has produced
recently very interesting and in some sense unexpected results
\cite{BV95,lathuile,BHV99,Berry,binger,fujii1,hannabuss,remarks,currents,fujii2,comment,bosonmix,Ji2,Ji3,3flavors}.
The story begins in 1995 when in Ref.\cite{BV95}, it was proved the
unitary inequivalence of the Hilbert spaces for (fermion)
fields with definite flavor
on one side and those (free fields) with definite mass, on the other. The
proof was then generalized to any number of fermion generations \cite{hannabuss}
and to bosonic fields \cite{lathuile,binger}.
This result strikes with the common sense of Quantum Mechanics (QM), where
one has only one Hilbert space at hand: the inconsistencies that arise there
have generated much controversy and it was also claimed that it
is impossible to construct an Hilbert space for flavor states \cite{kimbook}
(see however Ref.\cite{fujii1} for a criticism of  that argument).

In fact, not only the flavor Hilbert space can be consistently defined \cite{BV95},
but it also provide a tool for the calculation of flavor oscillation formulas in
QFT\cite{BHV99,remarks,currents,fujii2,comment,bosonmix,Ji2,Ji3,Beuthe,3flavors,BPT02,msw},
which exhibit corrections with respect to the usual QM ones \cite{Pontec,Cheng-Li}.

{}From a more general point of view, the above results show that mixing
is an ``example of non-perturbative physics
which can be exactly solved", as stated in Ref.\cite{Ji2}.
Indeed, the flavor Hilbert space is a space for particles which are  not
on-shell and this situation is analogous to that one encounters when quantizing
fields at finite temperature \cite{TFT} or in a curved background \cite{birrel}.

In the derivation of the oscillation formulas by use of the
flavor Hilbert space, both for bosons and for fermions,
a central role is played by the flavor charges \cite{currents} and indeed
it was found that these operators satisfy very specifical
physical requirements \cite{remarks,fujii1}.
However, these charges vanish identically in the case of neutral fields and
this is the main reason why the study of field mixing has been limited up to now
only to the case of charged (complex) fields.
We fill here this gap, by providing a consistent treatment of both neutral bosons
and Majorana fermions. In order to keep the discussion transparent,
we limit ourselves to the case of two generations.

Apart from the explicit quantization of the neutral mixed fields,
the main result of this paper is the study of the momentum  operator
(and more in general of the energy-momentum tensor)
for those fields. We show how to
define it in a consistent way and by its use we then derive the oscillation formulas,
which match the ones already obtained for charged fields.
We also comment on its relevance for the study of charged mixed fields,
where, in the case when CP violation is present, the charges present a problematic
interpretation which is still not completely clarified \cite{3flavors,fujii2,comment}.

The paper is organized as follow: in Section II we discuss the mixing  of neutral
(spin zero) bosonic fields. In Section III we treat Majorana
fields and finally in Section IV we discuss some  general consequences of our
results and draw conclusions.

\section{Neutral bosons}

We consider here the simple case of mixing of two spin zero
neutral boson fields. We follow the case of charged fields as discussed already in
Refs.\cite{lathuile,bosonmix,binger} and introduce the following lagrangian:
 \bea\label{boslagAB}
 {\cal L}(x)&=& \pa_\mu \Phi_f^T(x)
\pa^\mu \Phi_f (x)\, - \, \Phi_f^T(x) {\hat M}  \Phi_f(x)
\\[2mm]  \label{boslag12}
&=&\pa_\mu \Phi_m^T(x) \pa^\mu \Phi_m(x) \, - \, \Phi_m^T(x) {\hat
M}_d  \Phi_m (x) \eea
with $\Phi_f^T=(\phi_A,\phi_B)$ being the flavor fields and ${\hat M} =
\lf(\brr{cc} m_A^2 & m_{AB}^2 \\ m_{AB}^2 & m_B^2\err \ri)$. Those are
connected to the  free fields $\Phi_m^T=(\phi_1,\phi_2)$ with  ${\hat
M}_d = diag(m_1^2,m_2^2)$ by a rotation:
\bea\lab{2.53a}
&&\phi_{A}(x) = \phi_{1}(x) \; \cos\te + \phi_{2}(x) \; \sin\te
\\[2mm] \lab{2.53b}
&&\phi_{B}(x) =- \phi_{1}(x) \; \sin\te + \phi_{2}(x)\; \cos\te \eea
and a similar one for  the conjugate momenta $\pi_i=\pa_0\phi_i$.
The free fields $\phi_i$ can be quantized in the usual way (we use $x_0\equiv t$):
\bea\label{phimass}
\phi_{i}(x) &=& \int
\frac{d^3{\bf k}}{(2\pi)^{\frac{3}{2}}} \frac{1}{\sqrt{2\om_{k,i}}}
\lf( a_{{\bf k},i}\ e^{-i \om_{k,i} t } + a^{{\dag} }_{-{\bf k},i}\
e^{i \om_{k,i} t} \ri)e^{i {\bf k x}}
\\ [2mm] \label{pimass}
\pi_{i}(x) &=& -i\ \int \frac{d^3{\bf k}}{(2\pi)^{\frac{3}{2}}}
\sqrt{\frac{\om_{k,i}}{2}} \lf(a_{{\bf k},i}\ e^{-i \om_{k,i} t}
- a^{{\dag} }_{-{\bf k},i}\ e^{i \om_{k,i} t} \ri)e^{i {\bf k x}} \, , \quad i=1,2.
\eea
with $\om_{k,i}=\sqrt{{\bf k}^2 + m_i^2}$. The commutation relations are:
\bea\label{commutators}
\lf[ \phi_{i}(x),\pi_{j}(y)\ri]_{x_0=y_0}=
 i\, \de_{ij}\,\delta^{3}({\bf x}-{\bf y}) ~,~~~~
\lf[ a_{{\bf k},i}, a_{{\bf p},j}^{\dag} \ri]=
\, \de_{ij}\,\delta^{3}({\bf k}-{\bf p}).
\eea

We now recast Eqs.(\ref{2.53a}),(\ref{2.53b}) into the form:
\bea\lab{2.53c}
\phi_{A}(x) = G^{-1}_\te(t)\; \phi_{1}(x)\; G_\te(t) \\[2mm]
\lab{2.53d}
\phi_{B}(x) = G^{-1}_\te(t)\; \phi_{2}(x)\; G_\te(t)
\eea
and similar ones for $\pi_{A}(x)$, $\pi_B(x)$. $G_\te(t)$  denotes the
operator which implements the mixing transformations (\ref{2.53a}),(\ref{2.53b}):
\bea\lab{neutr} G_\te(t) = \exp\lf[-i\;\te \int d^{3}x
\lf(\pi_{1}(x)\phi_{2}(x)  -\pi_{2}(x)\phi_{1}(x) \ri)\ri]\, ,
\eea
which is (at finite volume) a unitary operator:
$G^{-1}_\te(t)=G_{-\te}(t)=G^{\dag}_\te(t)$.

The mixing 
generator is given by $G_{\te}(t) = \exp[\te(S_{+}(t) -
S_{-}(t))]$ and the $su(2)$ operators are now realized as
\bea \label{su2charges1}
&& S_{+}(t) \equiv -i\;\int d^{3}{\bf x} \; \pi_{1}(x)\phi_{2}(x)
\quad, \quad
S_{-} (t)\equiv -i\;\int d^{3}{\bf x} \;\pi_{2}(x)\phi_{1}(x)
\\ [2mm] \label{su2charges2}
&& S_{3} \equiv \frac{-i}{2} \int d^{3}{\bf x}
\lf(\pi_{1}(x)\phi_{1}(x) - \pi_{2}(x)\phi_{2}(x)\ri) \quad, \quad
S_{0} \equiv \frac{-i}{2} \int d^{3}{\bf x}
\lf( \pi_{1}(x)\phi_{1}(x)
+\pi_{2}(x)\phi_{2}(x) \ri)\, .
\eea
This has to be compared with the corresponding 
structures for charged fields \cite{bosonmix}.
We have, explicitly
\bea
&&S_{+}(t)-S_{-}(t)=\int d^{3}{\bf k}
\left( {\hat U}_{\bf k}^{*}(t) \; a_{{\bf k},1}^{\dag}a_{{\bf k},2}-
{\hat V}_{{\bf k}}^{*}(t) \; a_{-{\bf k},1}a_{{\bf k},2} + {\hat V}_{{\bf k}}(t) \;
a_{{\bf k},2}^{\dag}a_{-{\bf k},1}^{\dag}
- {\hat U}_{{\bf k}}(t) \; a_{{\bf k},2}^{\dag}a_{{\bf k},1} \right)
\eea
where ${\hat U}_{{\bf k}}(t)$ and ${\hat V}_{{\bf k}}(t)$
are Bogoliubov coefficients given by
\bea \lab{bog1}
&&{\hat U}_{{\bf k}}(t)\equiv |{\hat U}_{{\bf k}}| \; e^{i(\om_{k,2}-
\om_{k,1})t} \quad , \quad {\hat V}_{{\bf k}}(t)\equiv |{\hat V}_{{\bf k}}| \;
e^{i(\om_{k,1}+ \om_{k,2})t}
\\ [2mm]  \lab{bog2}
&&|{\hat U}_{{\bf k}}|\equiv \frac{1}{2}
\lf( \sqrt{\frac{\om_{k,1}}{\om_{k,2}}}
+ \sqrt{\frac{\om_{k,2}}{\om_{k,1}}}\ri) \quad , \quad
|{\hat V}_{{\bf k}}|\equiv  \frac{1}{2} \lf(
\sqrt{\frac{\om_{k,1}}{\om_{k,2}}}
- \sqrt{\frac{\om_{k,2}}{\om_{k,1}}} \ri)\quad, \quad
|{\hat U}_{{\bf k}}|^{2}-|{\hat V}_{{\bf k}}|^{2}=1\,,
\eea

The flavor fields can be expanded as:
\bea\label{phiflav}
\phi_{\si}(x) &=& \int
\frac{d^3{\bf k}}{(2\pi)^{\frac{3}{2}}} \frac{1}{\sqrt{2\om_{k,j}}}
\lf( a_{{\bf k},\si}(t)\ e^{-i \om_{k,j} t } + a^{{\dag} }_{-{\bf k},\si}(t)\
e^{i \om_{k,j} t} \ri)e^{i {\bf k x}}
\\ [2mm] \label{piflav}
\pi_{\si}(x) &=& -i\ \int \frac{d^3{\bf k}}{(2\pi)^{\frac{3}{2}}}
\sqrt{\frac{\om_{k,j}}{2}} \lf(a_{{\bf k},\si}(t)\ e^{-i \om_{k,j} t}
- a^{{\dag} }_{-{\bf k},\si}(t)\ e^{i \om_{k,j} t} \ri)e^{i {\bf k x}} \, , \quad
\eea
with $\si,j=(A,1),(B,2)$ and the flavor annihilation operators given by:
\bea
a_{{\bf k},A}(t)&\equiv&  G^{-1}_\te(t)\; a_{{\bf k},1}\; G_\te(t) \,=\,
\cos\te\;a_{{\bf k},1}\;+\;\sin\te\;
\lf( {\hat U}_{{\bf k}}^*(t)\;a_{{\bf k},2}\;+
\; {\hat V}_{{\bf k}}(t)\; a^{\dag}_{-{\bf k},2}\ri),
\\ \mlab{2.69}
a_{{\bf k},B}(t)&\equiv&  G^{-1}_\te(t)\; a_{{\bf k},2}\; G_\te(t) \,=\,
\cos\te\;a_{{\bf k},2}\;-\;\sin\te\;
\lf( {\hat U}_{{\bf k}}(t)\;a_{{\bf k},1}\;-
\; {\hat V}_{{\bf k}}(t)\; a^{\dag}_{-{\bf k},1}\ri).
\eea

We now consider the action of the generator of the
mixing transformations on the vacuum $|0 \ran_\mass$ for the
fields $\phi_{i}(x)$:  $a_{{\bf k},i}|0 \ran_\mass = 0, ~ i=1,2$ .
The generator induces an $SU(2)$ coherent state structure on such
state \cite{Per}:
\bea\label{2.61} |0(\te, t) \ran_\AB \equiv G^{-1}_\te(t)\; |0
\ran_\mass\,. \eea
{}From now on we will refer to the state $|0(\te, t) \ran_\AB$ as to
the {\em flavor vacuum} for bosons. The suffixes $A$ and $B$ label
the flavor charge content of the state. We have $\,_\AB\lan
0(\te,t)|0 (\te,t)\ran_\AB \,= \,1$. In the following, we will
consider the Hilbert space for flavor fields at a given time $t$,
say $t=0$, and it is useful to define $|0 (t)\ran_\AB\equiv|0(\te,
t)\ran_\AB$ and $|0 \ran_\AB\equiv|0(\te, t=0)\ran_\AB$ for future
reference. A crucial point is that the flavor and the mass vacua
are orthogonal in the infinite volume  limit. We indeed have:
\bea\label{bosort1} \,_\mass\lan 0|0 (t)\ran_\AB \,= \,
\prod\limits_{\bf k} \, _\mass\lan 0| G_{{\bf k},\te}^{-1}(t)
|0\ran_\mass\,=\,\prod\limits_{\bf k} \, \frac{1 }{1+\sin^{2}\te
|{\hat V}_{\bf k}|^{2}} \,\equiv\, \prod\limits_{\bf k} \,f_0^{\bf
k}(\te) ~, ~~~for ~ any ~ t, \eea
where we have used $G^{-1}_\te(t) = \prod\limits_{\bf k} \, G_{{\bf
k},\te}^{-1}(t)$ . In the infinite volume limit, we obtain
\bea\label{bosort2} \lim\limits_{V\rar \infty}\,_\mass\lan 0|0
(t)\ran_\AB = \lim\limits_{V\rar \infty}\, e^{\frac{V}{(2\pi)^3}
\int d^3 k \, \ln f_0^{\bf k}(\te) } \, = \, 0 ~, ~~~for ~ any ~
t. \eea
We have that $\ln f_0^{\bf
k}(\te)$ is indeed negative for any values of ${\bf k}$, $\te$ and
$m_1, m_2$ (note that $0\leq \te \leq \pi/4$). We also observe that
the orthogonality disappears when   $\te=0$ and/or $m_1=m_2$,
consistently with the fact that in both cases  there is no mixing.
We define the state for a mixed particle with ``flavor''
$A$ and momentum ${\bf k} $ as:
\bea
|  a_{{\bf k},A}(t)\ran &\equiv& a^{\dag}_{{\bf k},A}(t)|0 (t)\ran_\AB \, =\,
G^{-1}_\te(t)a^{\dag}_{{\bf k},1}|0\ran_\mass
\eea
In the following we work in the Heisenberg picture, flavor states will be taken at
reference time $t=0$ (including the flavor vacuum). We also define
$|a_{{\bf k},A}\ran\equiv |a_{{\bf k},A}(0)\ran$.

Let us now consider the (non-vanishing) commutators of the
flavor ladder operators at different times:
\bea \label{comma}
\Big[a_{{\bf k},A}(t), a^{\dag}_{{\bf k},A}(t')\Big]
&=& \cos^2 \te + \sin^2\te\,\lf( |{\hat U}_{\bf k}|^2 e^{-i (\om_{2} -\om_1) (t-t')}
- |{\hat V}_{\bf k}|^2 e^{i (\om_{2} +\om_1) (t-t')}\ri),
\\ [2mm] \label{commb}
\Big[ a^{\dag}_{-{\bf k},A}(t),a^{\dag}_{{\bf k},A}(t')\Big]
&=&\sin^2\te\,|{\hat U}_{\bf k}||{\hat V}_{\bf k}| \lf(e^{-i \om_{2}  (t-t')}-
e^{i \om_{2}(t-t') }\ri) e^{-i \om_{1}  (t+t')},
\\ [2mm]
\label{commc}
\Big[a_{{\bf k},B}(t), a^{\dag}_{{\bf k},A}(t')\Big]
&=& \cos\te\sin\te \, |{\hat U}_{\bf k}| \lf(e^{i (\om_{2} -\om_1) t'} -
 e^{i (\om_{2} -\om_1) t}\ri),
\\ [2mm] \label{commd}
\Big[a^{\dag}_{-{\bf k},B}(t), a^{\dag}_{{\bf k},A}(t')\Big]
&=&\cos\te\sin\te \, |{\hat V}_{\bf k}| \lf(e^{-i (\om_{2} +\om_1) t} -
 e^{-i (\om_{2} +\om_1) t'}\ri).
\eea

We observe that the following quantity is constant in time:
\bea\label{one}
\lf|\lf[a_{{\bf k},A}(t), a^{\dag}_{{\bf k},A}(t') \ri]\ri|^2
\, - \,\lf|\lf[a^{\dag}_{-{\bf k},A}(t), a^{\dag}_{{\bf k},A}(t') \ri]\ri|^2
\, + \,\lf|\lf[a_{{\bf k},B}(t),a^{\dag}_{{\bf k},A}(t') \ri]\ri|^2
\, - \, \lf|\lf[a^{\dag}_{-{\bf k},B}(t), a^{\dag}_{{\bf k},A}(t') \ri]\ri|^2
\, =\, 1\,.
\eea

In Ref.\cite{bosonmix}, the corresponding of Eq.(\ref{one}) for charged fields, was
consistently interpreted
as expressing the conservation of total charge\footnote{We have,
for charged bosonic fields:
\bea\non
 && \lan a_{{\bf k},A} |  Q_\si(t) |  a_{{\bf k},A}\ran\,
=\, \lf|\lf[ a_{{\bf k},\si}(t),  a^{\dag}_{{\bf
k},A}(0) \ri]\ri|^2 \; - \; \lf|\lf[b^{\dag}_{-{\bf
k},\si}(t), a^{\dag}_{{\bf k},A}(0) \ri]\ri|^2 \quad,
\qquad \si=A,B\, .
\eea
together with $\,_\AB\langle  0| Q_{\si}(t)|
 0\rangle_\AB =0$ and
$\sum_\si \lan
a_{{\bf k},A} |  Q_\si(t) |  a_{{\bf k},A}\ran \,
=\, 1$.}.
In the present case we are dealing with a neutral field
and thus the charge operator vanishes identically.
Nevertheless the quantities in Eq.(\ref{one})
are well defined and are the  neutral-field
counterparts of the corresponding ones for the case of charged fields.
Thus we look for a physical interpretation of such oscillating quantities.

Let us consider the momentum operator, defined as the diagonal space part
of the energy-momentum tensor\cite{Itz}:
$P^j\equiv\int d^{3}{\bf x}\Te^{0j}(x)$, with
$\Te^{\mu\nu}\equiv\pa^\mu\phi\pa^\nu\phi -
g^{\mu\nu}\lf[\frac{1}{2}(\pa\phi)^2-\frac{1}{2}m^2\phi^2\ri]$.
For the free fields $\phi_i$ we have:
\bea\label{pmass}
{\bf P}_i&=&\int d^{3}{\bf x} \;\pi_{i}(x){\bf \nabla}\phi_{i}(x)
\, =\, \int d^{3}{\bf k} \, \frac{\bf k}{2}\,
\left(a_{{\bf k},i}^{\dag}a_{{\bf k},i} \, -\,
 a_{-{\bf k},i}^{\dag}a_{-{\bf k},i} \ri) \quad,\quad i=1,2.
\eea
In a similar way we can define the momentum operator for mixed fields:
\bea\label{pflav}
{\bf P}_\si(t)&=&\int d^{3}{\bf x} \;\pi_{\si}(x){\bf \nabla}\phi_{\si}(x)
\, =\, \int d^{3}{\bf k} \, \frac{\bf k}{2}\,
\left(a_{{\bf k},\si}^{\dag}(t)a_{{\bf k},\si}(t) \, -\,
 a_{-{\bf k},\si}^{\dag}(t)a_{-{\bf k},\si}(t) \ri) \quad,\quad \si=A,B.
\eea
The two operators are obviously related:
$ {\bf P}_\si(t) =  G^{-1}_\te(t)\,{\bf P}_i\, G_\te(t)$. Note that the total momentum
is conserved in time since commutes with the generator of 
mixing transformations (at any time):
\bea
&&{\bf P}_A(t) \, + \, {\bf P}_B(t) \, =\,{\bf P}_1 \, + 
\, {\bf P}_2 \, \equiv\, {\bf P}
\quad, \quad \lf[{\bf P}\,, \,G_\te(t)\ri] \, =\, 0
\quad , \quad   \lf[ {\bf P}\,, \,H\ri] \, =\, 0\,.
\eea
Thus in the mixing of neutral fields, the momentum operator
plays an analogous r{\^o}le to that of
the charge for charged fields. For charged fields, the total
charge operator, associated with the
$U(1)$ invariance of the Lagrangian, is proportional to the
Casimir of the $SU(2)$ group associated
to the generators of the mixing transformations. This is not
true anymore for the case of neutral
fields, although the $SU(2)$ structure persists in this case as well
(see Eqs.(\ref{su2charges1}),(\ref{su2charges2})).

We now consider the expectation values of the momentum operator for flavor fields
on the flavor state $|a_{{\bf k},A}\ran$ with definite
momentum ${\bf k}$. Obviously, this  is an eigenstate of ${\bf P}_A(t)$
at time $t=0$:
\bea
{\bf P}_A(0) \,|a_{{\bf k},A}\ran \, =\, {\bf k}\,|a_{{\bf k},A}\ran\,,
\eea
which follows from ${\bf P}_1 \,|a_{{\bf k},1}\ran \, =\, {\bf k}\,|a_{{\bf k},1}\ran$ by
application of $G^{-1}_\te(0)$.
At time $t\neq 0$, the expectation value of the momentum 
(normalized to the initial value) gives:
\bea
{\cal P}_\si^A(t)&\equiv&
\frac{\lan a_{{\bf k},A} | {\bf P}_\si(t) | a_{{\bf k},A}\ran}
{\lan a_{{\bf k},A} | {\bf P}_\si(0) |  a_{{\bf k},A}\ran} \,
=\, \lf|\lf[ a_{{\bf k},\si}(t),  a^{\dag}_{{\bf k},A}(0) \ri]\ri|^2 \;
- \; \lf|\lf[a^{\dag}_{-{\bf k},\si}(t), a^{\dag}_{{\bf k},A}(0) \ri]\ri|^2 \quad,
\qquad \si=A,B\,
\eea
which is of the same form as the expression one obtains for the charged field.

One can explicitly check that
the (flavor) vacuum expectation value of the momentum operator ${\bf P}_\si(t)$ does
vanish at all times:
\bea
{}_\AB\lan 0 | {\bf P}_\si(t) | 0\ran_\AB \, =\, 0 \quad,
\qquad \si=A,B\,
\eea
which can be understood intuitively by realizing that
the flavor vacuum $| 0\ran_\AB$  does not carry momentum
since it is a condensate of
pairs carrying zero total momentum (like the BCS ground state, for example).

The explicit calculation of the oscillating quantities ${\cal P}_{{\bf k},\si}^A(t)$ gives:
\bea\label{Acharge}
{\cal P}_{{\bf k},A}^A(t)
&=& 1 - \sin^{2}( 2 \theta) \lf[ |{\hat U}_{{\bf k}}|^{2} \;
\sin^{2} \lf( \frac{\omega_{k,2} - \omega_{k,1}}{2} t \ri)
-|{\hat V}_{{\bf k}}|^{2} \;
\sin^{2} \lf( \frac{\omega_{k,2} + \omega_{k,1}}{2} t \ri) \ri]
\\ [4mm] \label{Bcharge}
{\cal P}_{{\bf k},B}^A(t)&=&
\sin^{2}( 2 \theta)\lf[ |{\hat U}_{{\bf k}}|^{2} \; \sin^{2} \lf(
\frac{\omega_{k,2} - \omega_{k,1}}{2} t \ri) -|{\hat V}_{{\bf k}}|^{2} \;
\sin^{2} \lf( \frac{\omega_{k,2} + \omega_{k,1}}{2} t \ri) \ri] \, .
\eea
in complete agreement with the charged field case\cite{bosonmix}. Note the
presence of the non-standard oscillation terms and of the momentum dependent
amplitudes. When $|{\hat U}_{{\bf k}}|^{2}\rar 1$ and 
$|{\hat V}_{{\bf k}}|^{2}\rar 0$ (see
Eq.(\ref{bog2})), the usual QM oscillation formulas are recovered.

\section{Majorana fermions}

Now we consider the case of mixing of two Majorana fermion fields. We closely follow
the above treatment for boson fields.
The charge-conjugation operator $\mathcal{C}$ is defined as
satisfying the relations
\be
\label{chargeconjoperator}
 {\cal C}^{-1}\gamma_\mu{\cal C} =
-\gamma_\mu^T \quad,\quad {\cal C}^{\dag} = {\cal C}^{-1}
\quad,\quad {\cal C}^{T} = -{\cal C} \ .
\ee
from which we define the charge conjugate $\psi^c$ of $\psi$ as
\be
\psi^c(x) \equiv \gamma_0{\cal C}\psi^*(x) \ .
\ee
Now we define a Majorana fermion as a field that satisfies the
Dirac equation
\be \label{eq:DiracEquation}
(i \!\not\!\partial - m) \psi = 0
\ee
and the self-conjugation relation
\be \label{eq:MajoranaCondition}
\psi = \psi^c \;.
\ee
Thus the two equations (\ref{eq:DiracEquation}) and
(\ref{eq:MajoranaCondition}) ensure that a
Majorana field is a  neutral fermion field.

We now proceed as in the above case for neutral bosons by introducing the
following Lagrangian:
\bea
{\cal L} (x) &=& \overline{\psi_f}(x) ( i\!\not\!\partial - M )
\psi_f(x)
\, =\,  \overline{\psi_m}(x) ( i\!\not\!\partial - M_d ) \psi_m(x)\,,
\eea
with $\psi_f^T = (\nu_e, \nu_\mu)$ being the flavor fields and
$M = \left( \brr{cc} m_e & m_{e\mu}
\\ m_{e\mu} & m_\mu \err \right)$. 
The flavor fields are connected to the free fields $\psi_m^T =
(\nu_1, \nu_2)$ with $M_d=diag(m_1, m_2)$ by the rotation:
\bea \label{eq:SimpleMajoranaMixing}
&&\nu_{e}(x) = \nu_{1}(x) \; \cos\te   + \nu_{2}(x) \; \sin\te\; ,
\\ [2mm] \label{eq:SimpleMajoranaMixing2}
 &&\nu_{\mu}(x) =- \nu_{1}(x) \; \sin\te   +
\nu_{2}(x)\; \cos\te\;.
\eea
The quantization of the free fields is given by \cite{Mohapatra}
\be
\nu_i (x) =
\sum_{r=1,2} \int \frac {d^3 {\bf k}} {(2\pi)^{\frac{3}{2}}}  e^{i{\bf
k}\cdot\mathbf{x}} \left[u^r_{{\bf k},i}(t)\al^r_{{\bf k},i} +  v^r_{-{\bf k},i}(t)
\al^{r{\dag}}_{-{\bf k},i}\right], \quad i = 1,2.\ee
where $u^r_{{\bf k},i}(t) = e^{-i\omega_{k,i}t}u^r_{{\bf k},i}$, $v^r_{{\bf k},i}(t) =
e^{i\omega_{k,i}t}v^r_{{\bf k},i}$, with $\omega_{k,i} = \sqrt{{\bf k}^2 + m^2_i}$. 
In order for
the  Majorana condition (\ref{eq:MajoranaCondition}) to be satisfied the four 
spinors must also
satisfy the following condition:
\bea \label{eq:MajoranaSpinorCondition}
v^s_{{\bf k},i} = \gamma_0 {\cal C} (u^{s}_{{\bf k},i})^*\,\quad;\quad
u^{s}_{{\bf k},i} = \gamma_0 {\cal C} (v^{s}_{{\bf k},i})^*\,.
\eea
The equal time anticommutation relations are:
\bea
\{\nu^{\al}_{i}(x), \nu^{\bt{\dag} }_{j}(y)\}_{t=t'} &=
\de^{3}{(\bf x-y)} \de_{\al\bt} \de_{ij} \quad,
\quad
\{\nu^{\al}_{i}(x), \nu^{\bt}_{j}(y)\}_{t=t'} &=
\de^{3}{(\bf x-y)} (\gamma_0 {\cal C})^{\al\bt}
\de_{ij} \;,
 \quad \al,\bt=1,..,4
\eea
and
\be
\{\al^{r}_{{\bf k},i}, \al^{s{\dag} }_{{\bf q},j}\} = \de^{3}{(\bf
k-q)}\de _{rs}\de_{ij}   \;,\qquad i,j=1,2\;.
\ee
All other anticommutators are zero. The orthonormality and
completeness relations are:
\be
u^{r{\dag}}_{{\bf k},i} u^{s}_{{\bf k},i} = v^{r{\dag}}_{{\bf k},i} 
v^{s}_{{\bf k},i} = \delta_{rs}
\;,\quad  u^{r{\dag}}_{{\bf k},i} v^{s}_{-{\bf k},i} = v^{r{\dag}}_{-{\bf k},i} 
u^{s}_{{\bf k},i} =
0\;,\quad \sum_{r=1,2}(u^{r}_{{\bf k},i} u^{r{\dag}}_{{\bf k},i} + v^{r}_{-{\bf k},i}
v^{r{\dag}}_{-{\bf k},i}) = I \;.
\ee

As for the neutral boson case we can recast
Eqs.(\ref{eq:SimpleMajoranaMixing}),(\ref{eq:SimpleMajoranaMixing2})
into the form:
\bea
&&\nu_e^{\al}(x) = G^{-1}_{\te}(t)\; \nu_{1}^{\al}(x)\; G_{\te}(t)\;,\\[2mm]
&&\nu_\mu^{\al}(x) = G^{-1}_{\te}(t)\; \nu_{2}^{\al}(x)\; G_{\te}(t)
\;,
\eea
where $G_{\te}(t)$ is given by
\be
G_{\te}(t) = \exp\lf[\frac {\te}{2} \int d^{3}{\bf x}  \lf(\nu_{1}^{\dag}(x) \nu_{2}(x) -
\nu_{2}^{\dag}(x) \nu_{1}(x) \ri)\ri]\;.
\ee
One has $G_{\te}(t)=\prod_{\bf k} G_{\te}^{\bf k}(t)$. For a given ${\bf k}$, in
the reference frame where ${\bf k}=(0,0,|{\bf k}|)$, the spins decouple \cite{BV95}
and one has
$G_{\te}^{\bf k}(t)=\prod_r G_{\te}^{{\bf k},r}(t)$ with
\be
G_{\te}^{{\bf k},r}(t) = \exp\lf\{\te  \left[
U_{\bf k}^*(t)\; \al^{r{\dag}}_{{\bf k},1} \al^{r}_{{\bf
k},2} - U_{\bf k}(t)\; \al^{r{\dag}}_{-{\bf k},2} \al^{r}_{-{\bf k},1} -
\ep^r V_{\bf k}^*(t) \al^{r}_{-{\bf k},1}
\al^{r}_{{\bf k},2} + \ep^r V_{\bf k}(t) \al^{r{\dag}}_{{\bf k},1}
\al^{r{\dag}}_{{-\bf k},2}\right]\ri\}\,,
\ee
where $U_{{\bf k}}(t)$ and $V_{{\bf k}}(t)$ are Bogoliubov coefficients given by
\be
U_{{\bf k}}(t)\equiv |U_{{\bf k}}|\;e^{i(\om_{k,2}-\om_{k,1})t} \quad,\quad
V_{{\bf k}}(t)\equiv|V_{{\bf
k}}|\;e^{i(\om_{k,2}+\om_{k,1})t}\;,
\ee
\bea
&&|U_{{\bf k}}|\equiv\lf(\frac{\om_{k,1}+m_{1}}{2\om_{k,1}}\ri)^{\frac{1}{2}}
\lf(\frac{\om_{k,2}+m_{2}}{2\om_{k,2}}\ri)^{\frac{1}{2}}
\lf(1+\frac{|\bf k|^{2}}{(\om_{k,1}+m_{1})(\om_{k,2}+m_{2})}\ri)\;,
\\[2mm]
 &&|V_{{\bf k}}|\equiv\lf(\frac{\om_{k,1}+m_{1}}{2\om_{k,1}}
\ri)^{\frac{1}{2}} \lf(\frac{\om_{k,2}+m_{2}}{2\om_{k,2}}\ri)^{\frac{1}{2}} \lf(\frac{|\bf
k|}{(\om_{k,2}+m_{2})}- \frac{|\bf k|}{(\om_{k,1}+m_{1})}\ri)\;,
\eea
\be
|U_{{\bf k}}|^{2}+|V_{{\bf k}}|^{2}=1 \;.
\ee

The flavor fields can be thus expanded as:
\be
\nu_\si (x) = \sum_{r=1,2} \int \frac {d^3 {\bf k}}
{(2\pi)^{\frac{3}{2}}} e^{i{\bf
k}\cdot\mathbf{x}} \left[u^r_{{\bf k},j}(t)\al^r_{{\bf k},\sigma}(t) +
v^r_{-{\bf k},j}(t)
\al^{r{\dag}}_{-{\bf k},\sigma}(t)\right], \ee
with $\sigma,j = (e,1),(\mu,2)$ and the flavor annihilation operators given by
(for ${\bf k}=(0,0,|{\bf k}|)$):
\bea
&&\al^{r}_{{\bf k},e}(t) \equiv G^{-1}_{\te}(t) \; \al^{r}_{{\bf k},1}\;
G_{\te}(t) = \cos \theta \,
\al^{r}_{{\bf k},1} + \sin \theta\,\lf(U_{\bf k}^*(t) \,\al^{r}_{{\bf k},2}+
\ep^r V_{\bf k}(t)\,
\al^{r{\dag}}_{{-\bf k},2}\ri)\;,
\\[2mm] \non
&&\al^{r}_{{\bf k},\mu}(t) \equiv G^{-1}_{\te}(t)\; \al^{r}_{{\bf k},2}\;
G_{\te}(t) = \cos \theta\,
\al^{r}_{{\bf k},2} -\sin \theta\, \lf(U_{\bf k}(t) \,\al^{r}_{{\bf k},1}
-\ep^r V_{\bf k}(t) \,\al^{r{\dag}}_{{-\bf k},1}\ri)\;.
\eea

As for the neutral boson case we now consider the action of the generator of the mixing
transformations on the vacuum $|0\rangle_\mass$. The flavor vacuum is defined as:
\be
|0(\theta,t)\rangle_{e,\mu} \equiv G_\theta^{-1}(t)|0\rangle_\mass\;.\
\ee

We define the state for a mixed particle with definite flavor, spin and momentum as:
\be
|\alpha^r_{{\bf k},e}(t)\rangle \equiv
\alpha^{r \dagger}_{{\bf k},e}(t)|0(t)\rangle_{e,\mu} =
G_\theta^{-1}(t) \alpha^{r \dagger}_{{\bf k},1}|0\rangle_\mass\;.
\ee

The anticommutators of the flavor ladder operators at different times are:
\bea
\left\{ \alpha^r_{{\bf k},e}(t),\alpha^{r \dagger}_{{\bf k},e}(t')
\right\} &=& \cos^2\theta +
\sin^2\theta \left( |U_{\bf k}|^2 e^{-i(\omega_2-\omega_1)(t-t')} + |V_{\bf
k}|^2 e^{i(\omega_2+\omega_1)(t-t')} \right) \;,
\\[2mm]
\left\{\alpha^{r \dagger}_{{-\bf k},e}(t),
\alpha^{r \dagger}_{{\bf k},e}(t')\right\}
&=& \epsilon^r\, \sin^2\theta \,|U_{\bf k}|
|V_{\bf k}|\, \left(e^{i\omega_2(t-t')} -
e^{-i\omega_2(t-t')}\right) e^{-i\omega_1 (t+t' )} \;,
\\[2mm]
\left\{ \alpha^r_{{\bf k},\mu}(t),\alpha^{r \dagger}_{{\bf k},e}(t') \right\} &=& \cos\theta
\sin\theta\,|U_{\bf k}| \,\left( e^{i(\omega_2-\omega_1)t'} -
e^{i(\omega_2-\omega_1)t} \right) \;,
\\[2mm]
\left\{\alpha^{r \dagger}_{{-\bf k},\mu}(t),\alpha^{r \dagger}_{{\bf k},e}(t')\right\}
&=& \epsilon^r\,
\cos\theta \sin\theta \, |V_{\bf k}| \,
\left( e^{-i(\omega_2+\omega_1)t'} - e^{-i(\omega_2+\omega_1)t}
\right) \;.
\eea
The following quantity is constant in time (notice the difference in the signs 
with respect to the boson case):
\be
\left|\left\{ \alpha^r_{{\bf k},e}(t),
\alpha^{r \dagger}_{{\bf k},e}(t') \right\}\right|^2 +
\left|\left\{\alpha^{r \dagger}_{{-\bf k},e}(t),
\alpha^{r \dagger}_{{\bf k},e}(t')\right\}\right|^2
+ \left|\left\{ \alpha^r_{{\bf k},\mu}(t),
\alpha^{r \dagger}_{{\bf k},e}(t') \right\}\right|^2 +
\left|\left\{\alpha^{r \dagger}_{{-\bf k},\mu}(t),
\alpha^{r \dagger}_{{\bf k},e}(t')\right\}\right|^2
= 1\;.
\ee

Again we consider the momentum operator defined as 
$P^j \equiv \int d^3{\bf x} {\cal J}^{0j}(x)$,
where the energy-momentum tensor for the fermion field, ${\cal J}^{\mu\nu}$, 
is defined by ${\cal
J}^{\mu\nu} \equiv i \overline{\psi}\gamma^\nu  \partial_\mu  \psi$. 
For the free fields $\psi_i$
we have:
\be
{\bf P}_i = \int d^3{\bf x}\, \psi^\dagger_i(x) (-i\nabla)\psi_i(x) =
\int d^3 {\bf k} \sum_{r=1,2} {\bf k}
 \left( \alpha^{r\dagger}_{{\bf k},i}\alpha^{r}_{{\bf k},i} - \alpha^{r\dagger}_{{-\bf
k},i}\alpha^{r}_{{-\bf k},i} \right)\quad , \quad i = 1,2 \;.
\ee
We then define the momentum operator for mixed fields:
\bea
{\bf P}_\sigma(t) = \int d^3{\bf x} \,\psi^\dagger_\sigma(x)
(-i\nabla)\psi_\sigma(x) = \int d^3 {\bf k}
\sum_{r=1,2} {\bf k} \left( \alpha^{r\dagger}_{{\bf k},\sigma}(t)
\alpha^{r}_{{\bf k},\sigma}(t) -
\alpha^{r\dagger}_{{-\bf k},\sigma}(t)\alpha^{r}_{{-\bf k},\sigma}(t)
\right)\quad , \quad \sigma = e,\mu
\;.
\eea
As for the boson case, we have ${\bf P}_\sigma(t)=
G_\theta^{-1}(t) {\bf P}_i G_\theta(t)$ and the conservation of 
total momentum as a consequence of
\bea
&& {\bf P}_e(t) + {\bf P}_\mu(t) = {\bf P}_1 + {\bf P}_2 \equiv {\bf P}\quad ,\quad
[{\bf P},G_\theta(t)] = 0 \quad , \quad [{\bf P}, H] = 0\;.
\eea

We now consider the expectation values on the flavor state
$|\alpha_{{\bf k},e}^r\rangle\equiv |\alpha_{{\bf k},e}^r(0)\rangle$.
At time $t=0$, this state is an eigenstate of the
momentum operator ${\bf P}_e(0)$:
\be
{\bf P}_e(0)\, |\alpha_{{\bf k},e}^r\rangle \, =\,
{\bf k} \,|\alpha_{{\bf k},e}^r\rangle \;.
\ee

At $t \neq 0$ the expectation value for the momentum (normalized to initial value) gives:
\be
{\cal P}^e_{{\bf k},\sigma}(t) \equiv \frac{\langle
\alpha^r_{{\bf k},e}|{\bf
P}_\sigma(t)|\alpha^r_{{\bf k},e}\rangle}
{\langle\alpha^r_{{\bf k},e}|{\bf
P}_\sigma(0)|\alpha^r_{{\bf k},e}\rangle} =
\left|\left\{ \alpha^r_{{\bf
k},\sigma}(t),\alpha^{r \dagger}_{{\bf k},e}(0) \right\}\right|^2 +
\left|\left\{\alpha^{r
\dagger}_{{-\bf k},\sigma}(t),\alpha^{r \dagger}_{{\bf k},e}(0)\right\}\right|^2
\quad , \quad
\sigma = e,\mu \;,
\ee
which is the same form of the expression one obtains for the expectation values of the
flavor charges in the case of Dirac  fields \cite{BHV99}.
The flavor vacuum expectation value of the momentum
operator ${\bf P}_\sigma(t)$ vanishes at all times:
\be
_{e,\mu}\langle 0|{\bf P}_\sigma(t)|0\rangle_{e,\mu} = 0 \quad , \quad \sigma = e,\mu \;.
\ee

The explicit calculation of the oscillating quantities 
${\cal P}^e_{{\bf k},\sigma}(t)$ gives:
\bea
{\cal P}^e_{{\bf k},e} (t) &=& 1 - \sin^2 2\theta \left[|U_{\bf k}|^2\,
\sin^2 \lf( \frac{\omega_{k,2} - \omega_{k,1}}{2} t \ri)  + |V_{\bf k}|^2\,
\sin^2\lf( \frac{\omega_{k,2} + \omega_{k,1}}{2} t \ri)\right]\;,
\\[2mm]
{\cal P}^e_{{\bf k},\mu} (t) &=& \sin^2 2\theta \left[|U_{\bf k}|^2\,
\sin^2 \lf( \frac{\omega_{k,2} - \omega_{k,1}}{2} t \ri)
+ |V_{\bf k}|^2\,
\sin^2 \lf( \frac{\omega_{k,2} + \omega_{k,1}}{2} t \ri)\right]\;,
\eea
in complete agreement with the Dirac  field case \cite{BHV99}.

\section{Discussion and Conclusions}

In this paper, we have studied in detail the mixing of neutral particles in the context
of Quantum Field Theory. We have considered explicitly the mixing among two generations,
both in the case of neutral scalar fields and for Majorana fields.
Our analysis confirm previous results on the mixing of charged fields,
and indeed we show  that also for neutral fields, the Hilbert spaces for definite flavor
and for definite masses are orthogonal in the infinite volume limit.

The main result of this paper is however the calculation of oscillation formulas, which
we obtained by use of the momentum operator, that is well defined for (mixed)
neutral fields, whereas the charge operator vanishes identically. Our results confirm
the oscillation formulas already obtained in the case of
charged fields by use of the flavor charges,
and it also reveals to be useful in the case of three flavor mixing, where the presence of
the CP violating phase introduces ambiguities in the treatment based on flavor
charges \cite{3flavors}.

It is indeed interesting to comment on this point:
for Dirac fields, the momentum operator is given as
\bea
{\bf P}_\sigma(t) =\,
\int d^3 {\bf k} \sum_r
\frac {\bf k} {2} \left( \alpha^{r\dagger}_{{\bf k},\sigma}(t)
\alpha^{r}_{{\bf k},\sigma}(t)  -
\alpha^{r\dagger}_{{-\bf k},\sigma}(t)\alpha^{r}_{{-\bf k},\sigma}(t) +
\beta^{r\dagger}_{{\bf k},\sigma}(t)\beta^{r}_{{\bf k},\sigma}(t)
-\beta^{r\dagger}_{-{\bf k},\sigma}(t)\beta^{r}_{-{\bf k},\sigma}(t)
\right)\quad , \quad \sigma = e,\mu,\tau\;.
\eea
This operator can be used for the calculation of the oscillation
formulas for Dirac neutrinos in analogy with what done above in
the Majorana case. Although in this case the charge operator is available and it
has been used successfully for deriving the oscillation formula \cite{BHV99} in the
two-flavor case, it has emerged that for three-flavor mixing,
the CP violating phase introduces
complications in the identification of the observables and indeed the matter is still
object of discussion \cite{3flavors,fujii2,comment}. The main problem there  is that the
flavor charges at time $t$ do not annihilate the flavor vacuum:
${}_f\lan 0 | Q_\si(t) | 0\ran_f \, \neq \, 0$ and this expectation value needs
to be subtracted by hand in order to get the correct oscillation formulas \cite{3flavors}.

However, we see easily how the use of the momentum operator confirms the results of
Ref.\cite{3flavors}, without presenting any ambiguity. We have indeed:
\bea
&&{}_f\lan 0 | {\bf P}_\si(t) | 0\ran_f\, =\, 0
\\ [2mm]
&&\frac{\langle \nu_\rho|{\bf P}_\si(t)| \nu_\rho\ran}
{\langle \nu_\rho|{\bf P}_\si(0)| \nu_\rho\ran}
 \;=\, \lf|\lf \{\al^{r}_{{\bf k},\si}(t), \al^{r
{\dag}}_{{\bf k},\rho}(0) \ri\}\ri|^{2} \;+ \;\lf|\lf\{\bt_{{-\bf
k},\si}^{r {\dag}}(t), \al^{r {\dag}}_{{\bf k},\rho}(0) \ri\}\ri|^{2}
\,,
\eea
with $\si,\rho = e,\mu,\tau$ and $|\nu_\rho\ran \equiv
\al^{r {\dag}}_{{\bf k},\rho}(0) |0\ran_f$.
This follows from the following relations:
\bea
&& {}_f\lan 0| \alpha^{r\dagger}_{{\bf k},\sigma}(t)
\alpha^{r}_{{\bf k},\sigma}(t)  | 0\ran_f \, =\,
{}_f\lan 0|\alpha^{r\dagger}_{{-\bf k},\sigma}(t)
\alpha^{r}_{{-\bf k},\sigma}(t) | 0\ran_f
\\ [2mm]
&& {}_f\lan 0| \bt^{r\dagger}_{{\bf k},\sigma}(t)
\bt^{r}_{{\bf k},\sigma}(t)  | 0\ran_f\, =\,
{}_f\lan 0|\bt^{r\dagger}_{{-\bf k},\sigma}(t)
\bt^{r}_{{-\bf k},\sigma}(t) | 0\ran_f
\eea
which are valid even in presence of CP violation, when
${}_f\lan 0| \alpha^{r\dagger}_{{\bf k},\sigma}(t)
\alpha^{r}_{{\bf k},\sigma}(t)  | 0\ran_f
\neq {}_f\lan 0| \bt^{r\dagger}_{-{\bf k},\sigma}(t)
\bt^{r}_{-{\bf k},\sigma}(t)  | 0\ran_f $.

These results seem to suggest that perhaps a redefinition of the
flavor charge operators is necessary in presence of CP violation and
further study in this direction is in progress.

\section*{Acknowledgements}

M.B. thanks the ESF network COSLAB and  EPSRC for support. J.P. thanks Oxford University for
support.


\begin{thebibliography}{99}
%
%
\bibitem{BV95}
M.~Blasone and G.~Vitiello,
Annals Phys.\  {\bf 244} (1995) 283 [Erratum-ibid.\  {\bf 249}
(1995) 363].


\bibitem{lathuile}
M.~Blasone, P.~A.~Henning and G.~Vitiello, in ``La Thuile 1996,
Results and perspectives in particle physics'' ed. M.Greco, INFN
Frascati 1996, p.139-152 [hep-ph/9605335].

\bibitem{BHV99}
M.~Blasone, P.~A.~Henning and G.~Vitiello,
Phys.\ Lett.\ B {\bf 451} (1999) 140;
M.~Blasone, in ``Erice 1998, From the Planck length to the Hubble
radius'' 584,
[hep-ph/9810329].

\bibitem{Berry}
M.~Blasone, P.~A.~Henning and G.~Vitiello,
Phys.\ Lett.\  B {\bf 466} (1999) 262 ;
X.~B.~Wang, L.~C.~Kwek, Y.~Liu and C.~H.~Oh,
Phys.\ Rev.\ D {\bf 63} (2001) 053003.

\bibitem{binger}
M.~Binger and C.~R.~Ji,
Phys.\ Rev.\ {\bf D60} (1999) 056005.


\bibitem{fujii1}
K.~Fujii, C.~Habe and T.~Yabuki,
Phys.\ Rev.\ D {\bf 59} (1999) 113003 [Erratum-ibid.\ D {\bf 60}
(1999) 099903].
;%


\bibitem{hannabuss}
K.~C.~Hannabuss and D.~C.~Latimer,
J.\ Phys.\ A {\bf 36} (2003) L69;
J.\ Phys.\ A{\bf A33} (2000) 1369.

\bibitem{remarks}
M.~Blasone and G.~Vitiello,
Phys.\ Rev.\ {\bf D60} (1999) 111302.

\bibitem{currents}
M.~Blasone, P.~Jizba and G.~Vitiello,
Phys.\ Lett.\ {\bf B 517 } (2001) 471.

\bibitem{fujii2}
K.~Fujii, C.~Habe and T.~Yabuki,
Phys.\ Rev.\ D {\bf 64} (2001) 013011.

\bibitem{comment}
M.~Blasone, A.~Capolupo and G.~Vitiello,
in *Zhang-Jia-Jie 2001, Flavor physics* 425-433
[hep-th/0107125];
M.~Blasone, A.~Capolupo and G.~Vitiello,
[hep-ph/0107183].


\bibitem{bosonmix}
M.~Blasone, A.~Capolupo, O.~Romei and G.~Vitiello,
Phys.\ Rev.\ D {\bf 63} (2001) 125015.


\bibitem{Ji2}
C.~R.~Ji and Y.~Mishchenko,
Phys.\ Rev.\ D 64 (2001) 076004.

\bibitem{Ji3}
C.~R.~Ji and Y.~Mishchenko,
Phys.\ Rev.\ D {\bf 65} (2002) 096015.



\bibitem{3flavors}
M.~Blasone, A.~Capolupo and G.~Vitiello,
Phys.\ Rev.\ D {\bf 66} (2002) 025033;


\bibitem{kimbook}
C.~Giunti, C.~W.~Kim and U.~W.~Lee,
Phys.\ Rev.\ D {\bf 45} (1992) 2414.
\\
C.~W.~Kim and A.~Pevsner,{\it Neutrinos in Physics and Astrophysics},
Harwood Ac. Press,  1993.



\bibitem{Beuthe}
M.~Beuthe,
Phys.\ Rev.\ D {\bf 66} (2002) 013003;
Phys.\ Rep.\  {\bf 375} (2003) 105.



\bibitem{BPT02}
M.~Blasone, P.~P.~Pacheco and H.~W.~C.~Tseung,
Phys.\ Rev.\ D {\bf 67} (2003) 073011.

\bibitem{msw}
K.~Fujii, C.~Habe and M.~Blasone,
[hep-ph/0212076].


\bibitem{Pontec}
S.~M.~Bilenky and B.~Pontecorvo, Phys.\ Rep.\ {\bf 41} (1978)
225.



\bibitem{Cheng-Li}
T.~Cheng and L.~Li, {\it Gauge Theory of Elementary Particle
Physics}, Clarendon Press, Oxford, 1989.


\bibitem{TFT}
H.~Umezawa, H.~Matsumoto and M.~Tachiki,
{\it Thermo Field Dynamics and Condensed States}, North-Holland Publ.Co., 
Amsterdam, 1982;
H.Umezawa,{\em Advanced Field Theory: Micro, Macro and Thermal
Physics} American Institute of Physics, 1993;
P.~A.~Henning,
Phys.\ Rept.\  {\bf 253} (1995) 235.

\bibitem{birrel}
N.~D. Birrell and P.~C.~W. Davies, {\it Quantum Field in Curved Space},
Cambridge University Press, 1982.


\bibitem{Per}
A.~Perelomov, {\it Generalized Coherent States and Their
Applications}, Springer--Verlag, Berlin, 1986.



\bibitem{Itz} C.~Itzykson and J.~B.~Zuber, {\it Quantum Field Theory}
McGraw-Hill Book Co., New York, 1980.

\bibitem{Mohapatra}
R.~Mohapatra and P.~Pal, {\it Massive Neutrinos in Physics and Astrophysics}, 
World Scientific, Singapore, 1998

\end{thebibliography}
\end{document}